# A set-up for Hard X-ray Time-resolved Resonant Inelastic X-ray Scattering at SwissFEL

Hui-Yuan Chen[1], Rolf B. Versteeg[1], Michele Puppin[1], Ludmila Leroy[1,2], Roman Mankowsky[3], Pirmin Böhler[3], Yunpei Deng[3], Linda Kerkhoff[4], Aldo Mozzanica[2], Roland Alexander Oggenfuss[3], Claude Pradervand[2], Mathias Sander[3], Grigory Smolentsev[5], Seraphin Vetter[2], Thierry Zamofing[3], Henrik T. Lemke[3], Majed Chergui[1,6*] and Giulia F. Mancini[1,7*]

[1] Lausanne Centre for Ultrafast Science (LACUS), ISIC, École Polytechnique Fédérale de Lausanne (EPFL), CH-1015 Lausanne, Switzerland

[2] Photon Science Division, Paul Scherrer Institut (PSI), 5232 Villigen, Switzerland

[3] SwissFEL, Paul Scherrer Institut (PSI), 5232 Villigen, Switzerland

[4] Sect. Crystallography, Institute of Geology and Mineralogy, University of Cologne, 50674 Kölln, Germany

[5] Energy and Environment Research Division, Paul Scherrer Institut (PSI), 5232 Villigen, Switzerland

[6] Elettra Sincrotrone, Strada Statale 14 - km 163,5, 34149 Basovizza, Trieste, Italy

[7] Laboratory for Ultrafast X-ray and Electron Microscopy (LUXEM), Department of Physics, University of Pavia, I-27100 Pavia, Italy.

* Corresponding authors: giuliafulvia.mancini@unipv.it; majed.chergui@epfl.ch

**Abstract :**

We present a new set up for resonant inelastic hard X-ray scattering at the Bernina beamline of SwissFEL with energy, momentum, and temporal resolution. The compact R=0.5 m Johann-type spectrometer can be equipped with up to 3 crystal analysers and allows efficient collection of RIXS spectra. Optical pumping for time-resolved studies can be realized with a broad span of optical wavelengths. We demonstrate the performance of the set-up at overall ~180 meV resolution in a study of ground-state and photoexcited (at 400 nm) honeycomb *5d* iridate α-$Li_2IrO_3$. Steady-state RIXS spectra at the Iridium $L_3$-edge (11.214 keV) have been collected and are in very good agreement with data collected at synchrotrons. The time-resolved RIXS transients (pumped minus unpumped spectra) exhibit changes in the energy-loss region <2 eV, whose features mostly result from the hopping nature of *5d* electrons in the honeycomb lattice. These changes are ascribed to modulations of the Ir-to-Ir intersite transition scattering efficiency, which we associate to a transient screening of the on-site Coulomb interaction.

## I. Introduction

The control of quantum phases is a focal point of modern condensed matter physics research. Through chemical doping, electric and magnetic fields or external pressure, critical points can be surpassed, leading to the formation of new phases of matter. An especially non-invasive control parameter is provided by laser pulses, permitting the tailoring of quantum materials properties through strong electric field effects and the generation of nonequilibrium quasiparticle densities [1–3]. Recent illustrative cases are ultrafast topology switching in Weyl semimetals [4], the electric-field enhancement of magnetic exchange interaction [5], metastable superconductivity [6], and the creation and control of light-induced magnetic vortex structures [7,8].

At the core of these complex phenomena is the interplay between non-equilibrium densities of the respective bosonic and fermionic quasiparticle densities belonging to the lattice, electronic, orbital and spin degree of freedom of a material. In order to disentangle the dynamics of the various subsystems, a complete snapshot of the nonequilibrium quasiparticle dispersions and distributions in momentum space is wished for. Here, optical techniques such as time-resolved Raman scattering provide sensitivity to both fermionic and bosonic quasiparticles but are ill-suited to provide large momentum-selectivity [9–11], whereas time- and angle-resolved photoemission spectroscopy (Tr-ARPES) [12–14] allows to capture fermionic quasiparticle dispersions and populations over a large momentum-range, but the bosonic degrees of freedom are indirectly accessed through coupling mechanisms [15,16]. The hiatus for large momentum selectivity probing of bosonic quasiparticles, i.e., phonons and magnons, as well as high energy resolution for charge-transfer, Mott-Hubbard gap, and crystal-field excitations, was recently bridged by time-resolved resonant inelastic X-ray scattering (tr-RIXS). RIXS probing relies on a second-order photon-in/photon-out process involving photoelectron transitions from core-levels to valence states. The energy loss between the incoming and outgoing photons and the corresponding momentum transfer permits the deduction of quasiparticle dispersion as in phonons [16,17], magnons [18,19], paramagnons [20], and charge-density waves [21].

With the advent of bright sources of ultrashort X-ray pulses, such as X-ray free electron laser (XFELs), the possibility opened up to extend RIXS spectroscopy into time-domain studies, suggesting a new probe for nonequilibrium light-induced phases and bosonic quasiparticle dynamics. Tr-RIXS in the soft X-ray region has unveiled photoinduced changes of charge order in high-$T_c$ cuprate superconductors [22,23], the orbital dynamics in laser-induced insulator-to-metal transition [24], the photoexcited kinetic process in the hematite photocatalyst [25], and the short-range magnetic correlation in charge-transfer insulators [26].

In heavy *4f* and *5d* compounds, exotic states of matter emerge as multiple electron correlations reach comparable magnitudes and intertwine, such as superconductivity [27–29], topological phases [30], and quantum spin liquid [31–34]. Only very recently were the first tr-RIXS experiments in the hard X-ray region accomplished at the Linac Coherent Light Source (LCLS) XFEL (Sanford, USA), unveiling magnon and spin dynamics in Mott insulators $Sr_2IrO_4$ [35] and antiferromagnet $Sr_3Ir_2O_7$ [36]. These pioneering experiments highlighted the potential of hard X-ray tr-RIXS to explore laser-driven phenomena in quantum materials with *5d* transition metals and *4f* rare earth metals [37]. Despite the rising interest in tr-RIXS experiments and the rapid theoretical progress in understanding nonequilibrium inelastic photon scattering [15,37–40], the number of available hard X-ray tr-RIXS apparatus around the world remains limited.

In this work, we present the hard X-ray tr-RIXS set-up at the beamline Bernina [48] of the SwissFEL [42,43] at the Paul Scherrer Institute (PSI), Villigen, Switzerland. Under an incident X-ray bandwidth of ~120 meV from a Si(333) monochromator at the Ir $L_3$-edge (11.214 keV), a Johann-type spectrometer with a Rowland radius of R=0.5 m provides an energy resolution of around ~180 meV and a spectroscopic resolving power approaching $10^5$. The commissioning experiment focuses on electronic dynamics in the honeycomb *5d* iridate α-$Li_2IrO_3$. We show that photoexcitation via the ligand-to-metal charge transfer (LMCT) state leads to a modulation in RIXS-scattering efficiency of the Ir-to-Ir inter-site excitation of α-$Li_2IrO_3$, which we tentatively attribute to a transient screening of the on-site Coulomb repulsion.

## II.     Instrumentation

Bernina is the condensed matter end station of the Aramis Hard X-ray arm at SwissFEL. [36] The hard X-ray flux reaches ~$5\times10^{13}$ photons/s at 11 keV ($3\times10^{14}$ photons/s at 2 keV) with a pulse length of 40 fs (FWHM) at 100 Hz repetition rate in high charge mode. The self-amplified spontaneous emission (SASE) hard X-ray source can be operated at photon energies of ~2 keV up to ~12.7 keV, spanning a wide range of metal *K* and *L* edges. For RIXS experiments, the SASE spectrum (~0.15% Bandwidth) is monochromatized using a Silicon double crystal monochromator (DCM) equipped with [311] and [111] oriented crystals. For experiments, which require higher energy resolution, Si(333) and (555) reflections can be used to reduce the bandwidth to ~120meV and ~26meV at the Ir $L_3$-edge (11.214 keV) which reduces transmissions to around $5\times10^{-4}$ and $1\times10^{-4}$, respectively. The monochromatized X-ray pulse is focused with a vertical spot size of 50 μm on the sample surface by a Kirkpatrick-Baez (KB) mirror, as shown in Fig. 1. A pulsed pump laser source tunable from UV to THz pulses may be used for sample excitation. The timing jitter between the optical pump and X-ray probe pulses is usually larger than the pulse duration, thus reducing the temporal resolution in time-resolved XFEL experiments. At the Bernina beamline, timing is measured and controlled by laser arrival monitors (LAM) and an XFEL bunch arrival monitor (BAM) synchronized to a master timing clock, yielding a temporal resolution of ~50 fs. Additional laser-to-X-ray arrival time measurements by cross-correlation allow data re-sorting and pulse-length limited time-resolution. Further details on the Bernina beamline at the SwissFEL can be found in Ref. [41].

The physical phenomena under investigation impose stringent criteria on the choice and design rationale of the spectrometer. Whereas for core-to-core level RIXS studies, the low energy and momentum resolving power of a von Hamos-type detector would suffice, collective modes in solid-state materials are only probed by core-to-valence RIXS and therefore require a larger Johann-type spectrometer, however at the expense of etendue. A Johann-type spectrometer was therefore selected for the tr-RIXS set-up at Bernina. In the Johann/Johansson type, scattered X-rays are collected by a spherically curved crystal (termed "analyzer") and focused on the detector in a point-to-point geometry. By scanning the incident Bragg angle on the analyzer while simultaneously positioning the detector on the Rowland circle, different emission energies fulfilling Bragg's law are recorded on the detector to obtain the energy loss spectrum. The radius of the Rowland circle, the analyzer radius and dicing, and the utilized Bragg reflection determine the final energy and momentum resolution.

The spectrometer design is inspired from those of the SuperXAS beamline of the Swiss Light Source (SLS, Villigen, Switzerland) and of beamline ID20 of the European Synchrotron Radiation Facility (ESRF, Grenoble, France) [44,45]. The combination of the essential components from both designs is adapted to accommodate the beamline machinery at Bernina for hard X-ray tr-RIXS experiments. The schematic instrumental configuration is depicted in figure 1. The incident hard X-ray beam first passes through a DCM, and is focused onto the sample surface at a grazing angle through the Kirkpatrick-Baez (KB) mirror. The scattered light from the sample is collected at ~90° with respect to the incidence beam direction to minimize elastic scattering. Inelastically-scattered X-rays are monochromatized by a diced spherical crystal analyzer (currently present: Si(533) and Si(884) with a dicing index of 1 mm x 1mm) of 100 mm diameter and 1m radius of curvature, which focuses the energy-dispersed X-rays onto the detector in a Rowland geometry of a radius of 0.5 m. Efficient hard X-ray detection is achieved with a JUNGFRAU 512 × 512 pixel charge-integrating hybrid pixel photon detector developed at PSI [46] with a pixel size of 75µm×75µm. This analyser configuration alone resolves 35 meV at the Ir $L_3$-edge using the herefore optimal Si(844) analyzers. Using the DCM at the Si(333) reflection and the Si(844) analyzer crystal, we estimate an overall energy resolution of 130 meV at the Ir $L_3$-edge, limited by the bandwidth of the incoming x-rays (see supplementary information S.2). Consequently, the energy resolution can further be enhanced to an estimated 60 meV using the Si(555) reflection of the DCM at the expense of a factor 5 reduction in the photon flux. The whole scattering beam path including analyzer and photon detector is enclosed inside a Helium-filled chamber in order to minimize the absorption and scattering of X-rays by air (Figure 2).

In the Johann scanning configuration, accurate and simultaneous positioning of the analyzer and detector is critical. The analyzer manipulator in Bernina is composed of four motorized translational and rotational stages, enabling fine positioning along all relevant spatial degrees of freedom. Two additional crystal analyzers next to the central analyzer, as shown in Figure 2, can be used to extend the effective collection solid angle for the most photon-hungry studies at the expense of momentum resolution and an increased alignment complexity [39,41]. Alternatively, the secondary analyzers can be used to simultaneously detect multiple Brillouin zones in crystalline samples [47]. The photon detector is mounted on a combination of two translational stages and one rotational stage to ensure accurate positioning of the photon-sensitive chip on the Rowland circle. The spectrometer design permits an easy mirroring of the detector arm configuration, by which the accessible momentum space is effectively doubled. The spectrometer is installed on a large theta/2-theta goniometer platform, which can carry an N-dimensional sample manipulator or a hexapod sample manipulator. Sample cooling can be realized by means of cryogenic blowers. A dedicated cryostat reaching a temperature of T=4 K is planned in the future.

### III. Experimental results and instrument performance

Strong spin-orbit coupled honeycomb materials have been suggested as a solid-state platform for the realization of a Kitaev spin liquid phase [48]. The edge-sharing geometry between the metal-ligand octahedra and the resulting dominant bond-directional exchange are prerequisites for the realization of this fractionalized spin state. However, the parasitic presence of isotropic interactions, stemming from small structural distortions away from the ideal honeycomb structure, prohibits the formation of a pure spin liquid phase at low temperatures [49]. Besides conventional tailoring of the magnetic state by chemical doping [50,51], pressure[47,48], or magnetic fields [52], various suggestions have been made to use light-matter interactions, more specifically, Floquet-engineering and photodoping [53–55], to drive materials towards a spin liquid phase. However, difficulties in probing the microscopic nature of the nonequilibrium magnetic state arise due to the apparent absence of long-range order. A suggestion is therefore to probe the transient electronic excitation spectrum in order to resolve electronic and exchange interactions in the photoinduced state [56,57]. The second-order nature of the RIXS process relaxes the selection rules while enabling larger energy transfer, covering a high momentum space, making it a preferred choice over optical probe techniques for this purpose.

In this experiment, we intend to evaluate the possibility of probing transient exchange interactions through time-resolved RIXS (tr-RIXS) spectroscopy in the honeycomb iridate α-$Li_2IrO_3$. In iridates, the large crystal field splitting leads over Hund's rule and results in all *5d* valence electrons occupying the lower $t_{2g}$ levels, leaving an empty $e_g$ level. The strong spin-orbit coupling further splits the $t_{2g}$ levels into a fully-filled *J=3/2* level and a half-filled *J=½* level. α-$Li_2IrO_3$ shows a monoclinic distortion away from the ideal honeycomb structure. This results in a slight mixing of the *J=½* level with higher-lying orbitals. In addition, the monoclinic distortion leads to isotropic exchange interactions, resulting in spiral antiferromagnetic spin order at low temperatures [49].

We first discuss the RIXS spectrometer performance under equilibrium conditions. For our study, a ~1×1 $mm^2$ single crystal α-$Li_2IrO_3$, grown by chemical transport reaction growth [58], was mounted on the goniometer stage. The experiment was performed at ambient conditions. The incident x-ray energy was tuned to the iridium $L_3$-edge at 11.214 eV. A 90-degree configuration suppresses the detection of elastically scattered X-ray photons. The Si(844) plane in a Si(533) diced analyzer was used to disperse the inelastically scattered light onto the Jungfrau detector. Figure 3 shows the recorded steady-state RIXS energy loss spectrum at k=[0 -9 3] and compares it to the measurement of the powder sample recorded by Gretarsson *et al.* at the MERIX spectrometer in 30 ID beamline of Advanced Photon Source (USA) [59,60]. The elastic peak has a FWHM of 177 ± 5 meV (Supplementary Figure S2), which provides us with an upper limit of the instrumental energy resolution for experiments at the Iridium $L_3$-edge. We note that this value is not far from the estimated resolution (128 meV) bearing in mind the fact that it is affected by the phonon broadening at room temperature and the skewed Bragg reflection from the Si(844) plane of the Si(533) diced analyzer (Supplementary Information S2).

Compared to Gretarsson *et al.*, despite our lower energy resolution primarily due to a smaller Rowland circle, the measured RIXS spectrum captures all the features observed in the synchrotron study. The various exchange interactions and non-trivial crystallographic structure of α-Li$_2$IrO$_3$ lead to a rather intricate RIXS spectrum composed of various on-site and inter-site electronic excitations. In figure 3(a), on-site spin-orbit (SO) excitations ($J_{3/2}$ -> $J_{1/2}$) are found around 0.7~1.0 eV. The experimental resolution does not permit further resolving this localized excitation manifold [17,59]. The broad energy loss peak around 3-4 eV consists of $t_{2g}$-$e_g$ orbital *d-d* excitations, which are also localized. More important to our study are the delocalized excitations around energy losses of 0.4 eV, 1.3 eV, and 1.6 eV, as depicted in figure 3(b), which result from the hopping nature of *5d* electrons in the honeycomb lattice. This assignment is based on theoretical and experimental studies. Kim *et al.* [61] showed that these two features are absent in the single-site simulation but emerge in a four-site cluster simulation, which points to their itinerant origin. Further experimental evidence from high-resolution oxygen (O) *K*-edge RIXS [17] revealed that the feature near 0.4 eV is a dispersive spin-orbit exciton, whereas the features of 1.3 eV and 1.6 eV are inter-site excitations between the spin-orbit coupled $t_{2g}$ levels of two neighboring Ir sites in the honeycomb, which also aligns with optical conductivity studies [62,63]. The energy of 0.4 eV lies right across the Mott gap of honeycomb iridates [62], and we believe it represents the low energy excitation of "bound" holon-doublon pairs, i.e., Hubbard excitons [64,65], with a dressed mixing of local spin-orbit excitons, whereas 1.3 eV provides more energy to create "unbound" holons and doublons. Since the *L*-edge RIXS predominately probes the local environment, the appearance of these two itinerant features hints to a clear degree of mixing between inter-site and on-site excitation.

For the time-resolved experiment, we photoexcite the α-Li$_2$IrO$_3$ sample with 400 nm (3.1 eV) optical pulses at a fluence of ~45 mJ/cm$^2$. RIXS spectra over 0 to 2 eV energy loss range were recorded at -50 ps before the optical pump pulse and +1 ps after it, corresponding to an unpumped and pumped spectrum, respectively. Based on the electronic structure calculation [66] and the photoemission experiment [62], a pump energy of 3.1 eV corresponds to an ligand-to-metal charge-transfer (LMCT) dipole excitation from the oxygen *2p* orbital to the Ir atoms. Figure 4(a) shows the unpumped (static) RIXS spectrum (black solid circle) versus the spectrum at 1 ps after photoexcitation (purple circle) in the 0 to 2 eV energy loss region, while the lower panel shows the transient RIXS spectrum calculated by (RIXS$_{pumped}$-RIXS$_{unpumped}$)/RIXS$_{unpumped}$. Zooms to two specific energy loss regions are given in Figures 4b and 4c. These results were collected over 8 hours of data acquisition at a repetition rate of 100 Hz and a photon flux of around $10^{11}$ photons/s of the monochromatic X-ray.

Although the transient RIXS spectrum is quite noisy, we note the following: I) two features that overshoot the variation of the elastic peak (grey-shaded area) appear at ~0.4 eV and 1.3-1.6 eV, corresponding to the itinerant behaviour of the honeycomb *5d* electrons, and are zoomed into in Figures 4(b) and 4(c), respectively; II) although the 0.7 eV feature is the most intense in the steady-state RIXS spectrum as a localized excitation, there is no clear hint of a response there; c) the signal observed at ~0.4 eV is relatively small compared to the signal in the 1.3-1.6 eV region. Five Lorentzian peaks were fitted to both the pumped and unpumped RIXS spectra separately (Supplementary Information S3), which include: (i) the elastic scattering line; (ii) the spin-orbit exciton ~0.4 eV; (iii) the on-site $J_{3/2}$ to $J_{1/2}$ transition 0.7 eV; (vi) the inter-site $J_{1/2}$ to $J_{1/2}$ transition ~1.3 eV, and; (v) the inter-site $J_{3/2}$ to $J_{1/2}$ transition ~1.6 eV. The fit results are shown in Supplementary Figure S3. The orange trace in Figure 4a compares the transient tr-RIXS spectrum with the transient one calculated from the fitted spectra in Figures S3(a) and S3(b). They clearly show a much weaker 0.4 eV feature in the difference spectrum compared to the 1.3 and 1.6 eV features. Regarding the latter two, the fitted 1.3 eV peak undergoes an intensity decrease, while minimal change occurs to the fitted peak in the 1.6 eV region.

Tailoring the exchange interaction in iridate honeycomb materials, more specifically through the modulation of the Ir-site covalency, has been demonstrated by metal intercalation [56]. A similar mechanism may lie at the origin of the change in the inter-site excitation scattering rate in α-Li$_2$IrO$_3$. Let us first consider the possible inter-site excitation picture in the RIXS process without optical photoexcitation. Since it involves two sites, the initial state can be written as *5d$_i^5$5d$_j^5$*, where *i, j* denote two neighboring Ir sites. X-ray *L*-edge absorption creates an intermediate state: *2p5d$_i^6$5d$_j^5$*, where *2p* denotes the Ir *2p* core hole. Lastly, via recombination of core hole with the electron from the neighboring site, the system reaches a final state: *5d$_i^6$5d$_j^4$*, leaving an inter-site excitation from site j to site i compared to the initial state. When the system is pumped, the strong CT photoexcitation induces a significant disequilibrated density of holes on the oxygen site and doubly occupied Ir-sites (doublons, as *5d$^6$*). The initial state for the RIXS process has changed due to ligand holes and doublon population, which can be expressed as: α|*5d$_i^5$;5d$_j^6$L*> + β|*5d$_i^6$L;5d$_j^5$*> + γ|*5d$_i^6$L;5d$_j^6$L* > , where *L* denotes the O 2p ligand hole and $α^2+β^2+γ^2=1$. The induced screening of the on-site Coulomb repulsion U alters the Ir-O-Ir superexchange pathway strength set by ~$t^2$/U, with *t* being the pathway's hopping parameter [67]. The far-from-equilibrium electronic distribution leads to a change in the energy necessary for the Ir-Ir inter-site hopping, observed as a modulation of the RIXS matrix element. Any pump-induced effect on the quasiparticle population is only expected for the lowest energy transfer region <0.1eV, i.e., the asymmetric Stokes and anti-Stokes wings [9,10,34], which is below the resolution limit of our experiment. Similar modulation of the exchange interaction with ultrafast laser pulses has also been reported in other systems, such as iron oxides [68] and ferromagnetic insulator $CrSiTe_3$ [69]. A more detailed discussion of our results will be presented in a separate publication.

### IV. Summary and outlook

In summary, we presented the tr-RIXS set-up at the hard X-ray Bernina beamline at SwissFEL (Paul Scherrer Institute, Villigen, Switzerland). The compact R=0.5 m Johann-type spectrometer, which can be equipped with up to 3 crystal analyzers, permits efficient collection of RIXS spectra with a resolution of ~150 meV. Pumping can be realized with a broad span of optical wavelengths. We demonstrated the functioning of the set-up by photoexciting the honeycomb iridate α-$Li_2IrO_3$ in the charge transfer manifold. Modulation of the Ir-to-Ir intersite transition scattering efficiency was observed, which we ascribe to a transient screening of the on-site Coulomb interaction.

We note that magnetism and spin dynamics research could benefit from higher energy resolution (~50 meV) of the analyser and the monochromator to resolve low-energy modes. Nevertheless, the hard X-ray tr-RIXS set-up described here is readily capable of dynamical studies in the fields of chemistry, biology, and catalysis. An energy resolution of ~150 meV is sufficient to resolve plasmons and element-specific charge-transfer excitations within the local chemical environment [70], in addition to a threefold-extended solid angle to collect scattering in liquid/gas phase environment. Using hard X-ray tr-RIXS, real-time tracking of oxygen evolution reaction (OER) in iridium oxide [71], and CO adsorption dynamics in Pt nanoparticles (Pt $L_3$-edg~11.5 keV) could be performed [72]. We believe the newly commissioned hard X-ray tr-RIXS set-up at Bernina, SwissFEL, will enable unique research in laser-driven quantum materials and ultrafast X-ray spectroscopy.

**Acknowledgements**

This project was funded by the ERC Advanced Grant DYNAMOX (ID 695197). We thank Marco Moretti Sala and colleagues for discussions on the design of the Johann-type RIXS spectrometer. We thank Markus Grüninger and Alessandro Revelli for exchanging information on RIXS spectroscopy and α-$Li_2IrO_3$ during the commissioning experiment. We thank Markus Grüninger and Bo Yuan for the insightful discussion on the results. We acknowledge the Paul Scherrer Institute, Villigen, Switzerland for provision of free-electron laser beamtime at the Bernina instrument of the SwissFEL ARAMIS branch. MC acknowledges support from the ERC Advanced Grant CHIRAX (ID 101095012). G.F.M acknowledges support from ERC Starting Grant ULTRAIMAGE (851154), ERC PoC HYPER (101123123), Fondazione Cariplo 2020-2544 (NANOFAST), MUR PiXiE (R207A8MNNJ) and PRIN2022 DynaMAT (2022PR7CCY).

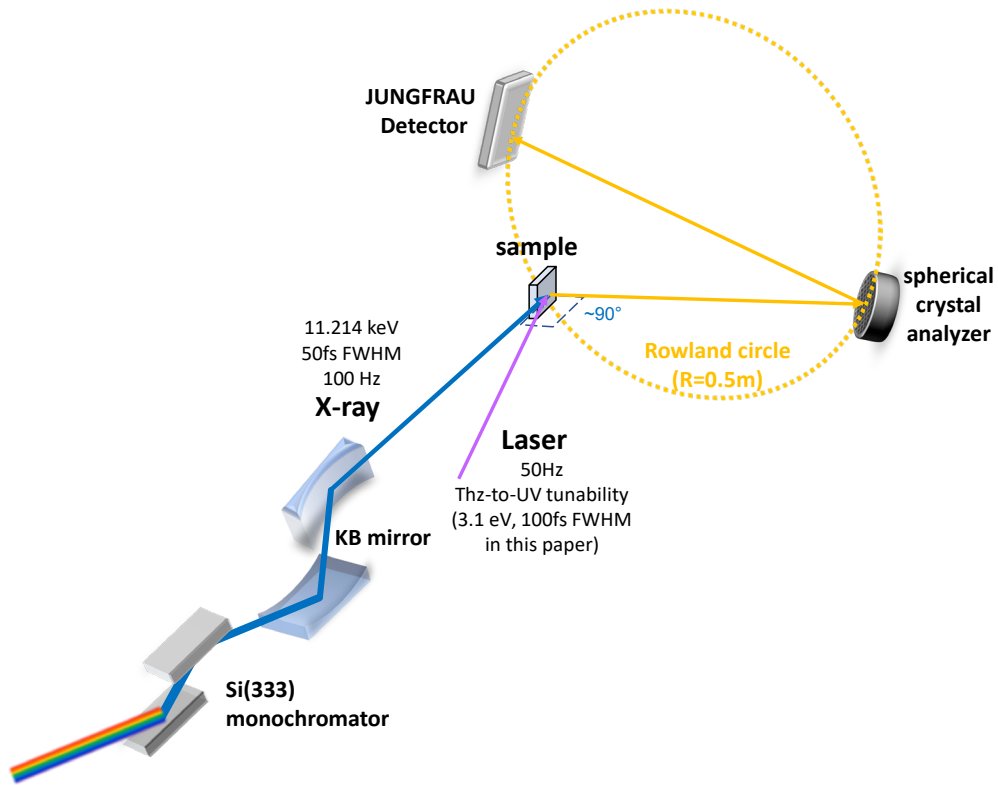

Figure 1. Overview of the tr-RIXS set-up at the Bernina beamline of SwissFEL. The incident X-ray is tuned to the Ir $L_3$-edge at 11.214 keV, after passing through an Si(333) monochromator before being focused onto the sample. The yellow dashed line represents the Rowland geometric relationship between the sample, analyzer, and detector.

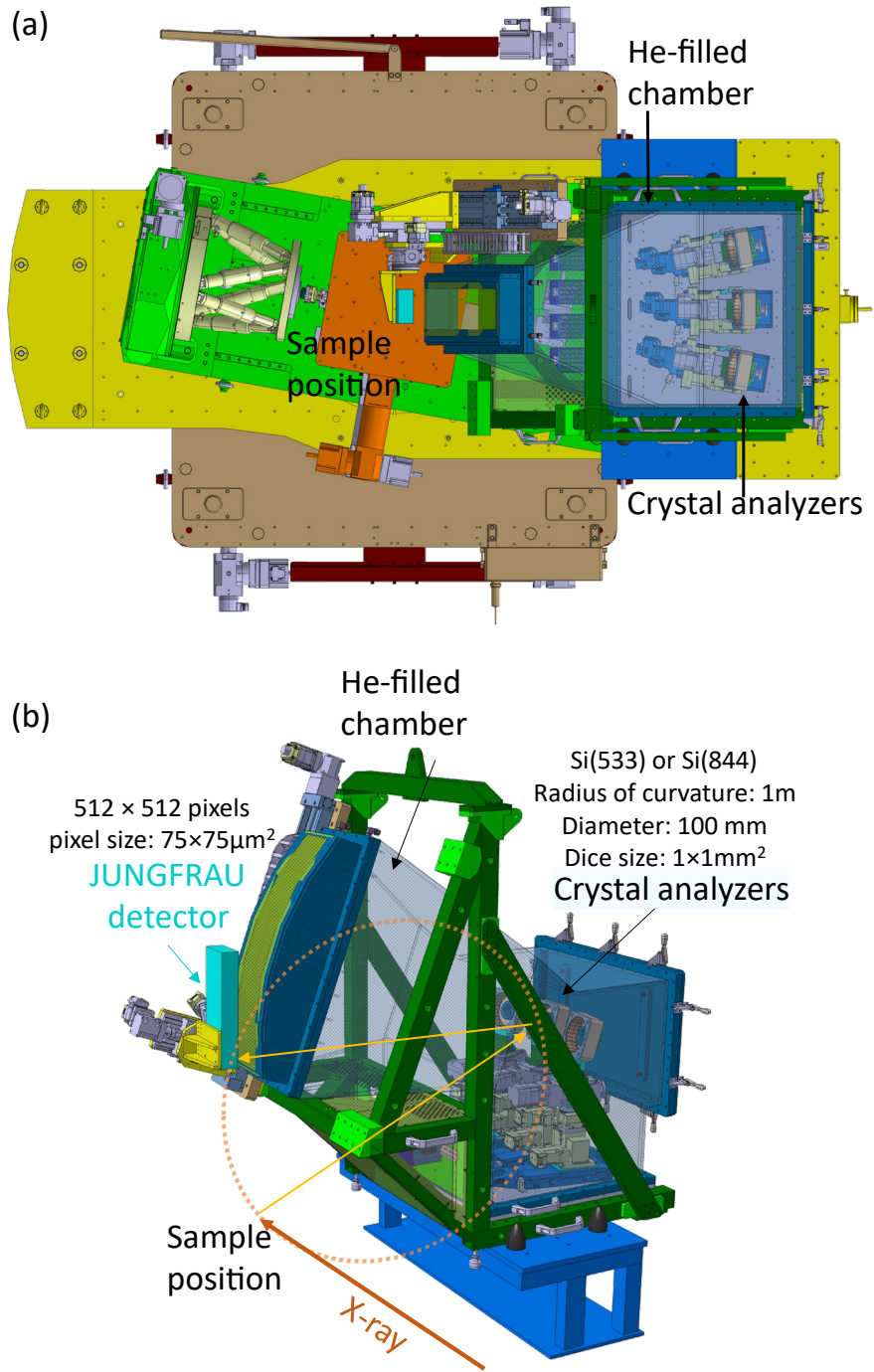

Figure 2. The Johann-type spectrometer at the Bernina beamline of SwissFEL. (a) Top view of the set-up, including the sample manipulator. The spectrometer is mounted on a general-purpose station. (b) Side view of the spectrometer alone, which comprises three crystal analyzers and the JUNGFRAU detector. The yellow dotted circle indicates the Rowland geometrical relationship between the sample, detector, and principal analyzer in the middle.

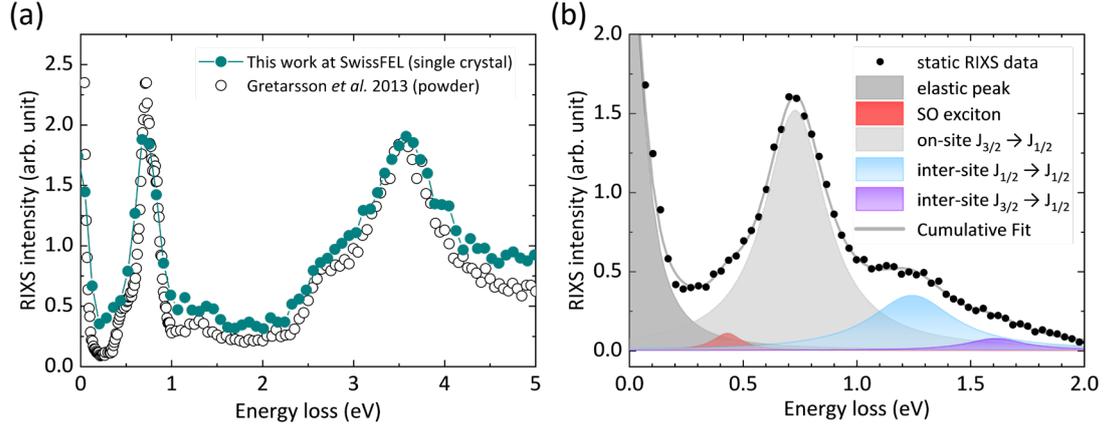

Figure 3. (a) Comparison of the Ir $L_3$-edge (11.214 keV) static RIXS spectrum recorded here and at the ESRF by Gretarsson et al. [59]. (b) Representation of localized and delocalized features as described in the main text.

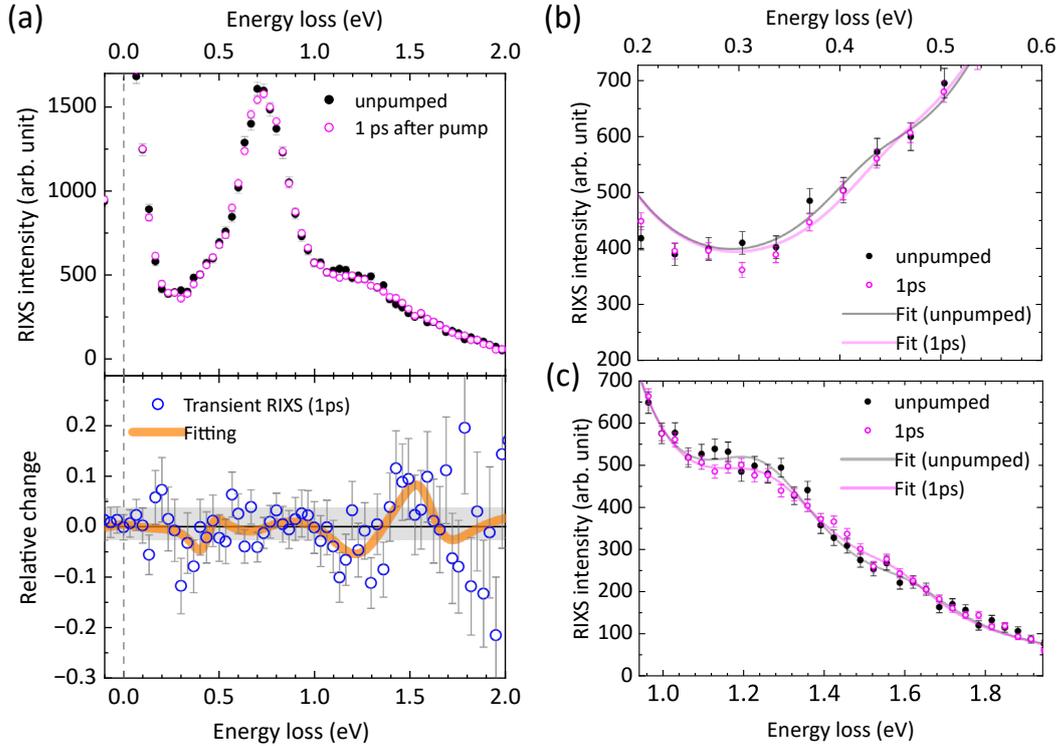

Figure 4. (a) Upper panel: Ir $L_3$-edge tr-RIXS spectra of α-Li$_2$IrO$_3$ before (black) and 1 ps after (magenta) excitation by a femtosecond pulse at 400 nm. Lower panel: Transient RIXS spectrum at 1 ps time delay (blue circles) in relative change with respect to the steady-state spectrum. The large scatter of the data points at high energy loss spectrum is due to the division by the low signal in the steady-state spectrum. The solid orange line indicates the differential from fitting with 5 Lorentzian peaks (see Supplementary information S3). The dashed line indicates the elastic line and the grey shaded area represents the variation of elastic peak. (b) Zoom into the energy loss spectrum around 0.4 eV. (c) Zoom into the energy loss spectrum around 1.3 eV. The error bars are estimated by Poisson statistics and error propagation.

# Supplementary Information

## A set-up for Hard X-ray Time-resolved Resonant Inelastic X-ray Scattering at SwissFEL

Hui-Yuan Chen[1], Rolf B. Versteeg[1], Michele Puppin[1], Ludmila Leroy[1,2], Roman Mankowsky[3], Pirmin Böhler[3], Yunpei Deng[3], Linda Kerkhoff[4], Aldo Mozzanica[2], Roland Alexander Oggenfuss[3], Claude Pradervand[2], Mathias Sander[3], Grigory Smolentsev[5], Seraphin Vetter[2], Thierry Zamofing[3], Henrik T. Lemke[3], Majed Chergui[1,6\*] and Giulia F. Mancini[1,7\*]

[1] Lausanne Centre for Ultrafast Science (LACUS), ISIC, École Polytechnique Fédérale de Lausanne (EPFL), CH-1015 Lausanne, Switzerland

[2] Photon Science Division, Paul Scherrer Institut (PSI), 5232 Villigen, Switzerland

[3] SwissFEL, Paul Scherrer Institut (PSI), 5232 Villigen, Switzerland

[4] Sect. Crystallography, Institute of Geology and Mineralogy, University of Cologne, 50674 Kölln, Germany

[5] Energy and Environment Research Division, Paul Scherrer Institut (PSI), 5232 Villigen, Switzerland

[6] Elettra Sincrotrone, Strada Statale 14 - km 163,5, 34149 Basovizza, Trieste, Italy

[7] Laboratory for Ultrafast X-ray and Electron Microscopy (LUXEM), Department of Physics, University of Pavia, I-27100 Pavia, Italy.

\* Corresponding authors: giuliafulvia.mancini@unipv.it; majed.chergui@epfl.ch

### S.1 $L_3$-edge X-ray absorption spectrum

The X-ray absorption spectrum (XAS) of our sample in the region of the $L_3$ edge of Iridium [1,2] is shown in figure S1. It was recorded at the same beamline (Bernina) where the RIXS measurements were carried out. Since we study the spin-orbit coupling and the pseudo-spins, the excitation energy of the RIXS was tuned to be at resonance with the $t_{2g}$ level, which corresponds to an energy of 11.214 keV according to Revelli *et al.* [2].

### S.2 Energy resolution

The energy resolution of a RIXS spectrometer can be approximated as the square sum of bandwidth contribution from all resolution elements in the Rowland geometry [3],

$$\Delta E_{total} = \sqrt{\Delta E_s^{\,2} + \Delta E_a^{\,2} + \Delta E_d^{\,2} + \Delta E_i^2}$$

where:

- $\Delta E_s$ = 41.97 meV is the bandwidth contribution for a spot size of 50 μm at the Bragg angle of 85.72° at 1 m;

- $\Delta E_a$ = 14.6 meV is the bandwidth of the Si(844) analyzer [4];

- $\Delta E_d$ = 31.48 meV is the detector contribution for a pixel size of 75 μm at the Bragg angle of 85.72° at 1 m;

- $\Delta E_i$ = 116.11 meV is the incidence bandwidth at the Bragg angle of 31.93° of the Si(333) monochromator.

Hence, the total energy bandwidth $\Delta E_{total}$ is estimated to be 128.24 meV. Using an incident X-ray bandwidth of 27 meV as calculated for the Si(555) reflection of the DCM and focusing the beam vertically to 10 um, we estimate a total energy resolution of $\Delta E_{total} \approx 45$ meV. The 1 mm sized dices cover 0.84 eV without scanning the energy of the RIXS analyzer.

The experimental energy resolution of 177 (meV) is evaluated by fitting a Voigt profile to the elastic peak measured at 300 K, shown in figure S2. The difference between the calculated and the measured value is attributed to phonon broadening at room temperature and the skewed Bragg reflection from the Si(533) analyzer instead of the right cut of Si(844) one. It is due to a logistical issue during the beamtime. As the Si(844) analyzer was unavailable, we adopted a Si(533)-cut crystal analyzer and tilted it to use the Si(844) surface. Currently the Bernina beamline is equipped with both analyzers.

### *S.3 Fits of the tr-RIXS spectra*

According to the experimental studies of the optical conductivity [5] and the high-resolution oxygen *K*-edge RIXS [6], there are four prominent excitation features around the spin-orbital coupling region from 0 to 2 eV energy loss, we therefore used 5 Lorentzian peaks to fit the spectra: Peak 1=elastic peak; Peak 2 is the spin-orbit exciton at ~0.4 eV; Peak 3 is the on-site $J_{3/2}$ to $J_{1/2}$ transition at ~0.7 eV; Peak 4 is the inter-site $J_{1/2}$ to $J_{1/2}$ transition at ~1.25 eV, and; Peak 5 is the inter-site $J_{3/2}$ to $J_{1/2}$ transition at ~1.6 eV.

We first fit the unpumped ground-state spectrum ("pump_off") as shown in figure S3(a), and then fit the pumped spectrum using the fit parameters obtained from the ground-state spectrum as starting values. We fixed the Lorentz area parameter (A), while allowing other parameters such as peak position and linewidth to change. Figure S3(b) shows the fitted pumped RIXS spectra ("pump_on"). All fitted parameters and corresponding standard errors are reported in Table S1 and S2.

In figure S4, we present the transient tr-RIXS spectra together with fitted peak for better visibility. The transient tr-RIXS spectra is defined in relative change as:

$$\frac{RIXS_{on} - RIXS_{off}}{RIXS_{off}}$$

Figures S4 (a) and (d) show all the fitted peaks, while figures S4 (b) and (e) are zooms-in of the 0.4 eV region, and figure S4(c) and (f) are zooms-in of the 1.3~1.6 eV region. The grey areas in figure S4 (a), (b), (c) indicate the range encompassed by the error and fluctuation of the elastic peak (which we assumed to be unchanged after the pump), serving as an estimate of the uncertainty range for other fitted peak changes.

However, it is challenging to quantify the modulation of the fitted peak after pump excitation. In Figure S5, we also show another fit of the pumped spectrum without the constraint on the area parameters, where the fitted result gives the same quality of fit (R-square =0.987) as the one presented in the main text and in Figure S3. Zooming into the itinerant excitation regions (0.4 eV and 1.6 eV) of the pumped spectrum, Figure S5 compares these two fit results. They show distinct results on the modulation of Peaks 2, 4, and 5, which is a consequence of the proximity of Peak 4 and 5 and the fact that all three peaks are extremely broad. This demonstrates how challenging it is to quantify the pump's impact on the peaks in terms of shifting, broadening, or intensity modulation with the current energy resolution presented. Nevertheless, there is no doubt that the peaks are qualitatively perturbed despite the difficulty in quantifying how much.

On the other hand, it might seem that the fit of peak 2 (SO exciton ~0.4 eV) is superfluous since the change is small and the two fits presented in Figure S5 give completely different results: one shows shifting (Figure S5b) while the other shows an intensity decrease (Figure S5e), making one question the necessity of fitting this peak. However, we show that, even though the exact modulation (shifting/broadening/decreasing) cannot be quantified, the inclusion of peak 2 indeed captures better the transient difference spectrum around 0.4 eV. Figure S6 plots the difference spectra (pumped subtracts unpumped) and compares the fit results among different inclusion of the Peaks to be fitted. It is shown that an improved reproduction of the transient data when all the excitation features (Peak 1 to 5) reported in the literature are included.

Table S1. Peak fitting of unpumped spectrum

| Model | Lorentz | | | | |
|---|---|---|---|---|---|
| Equation | y = y0 + (2*A/pi)*(w/(4*(x-xc)^2 + w^2)) | | | | |
| Plot | Peak1(pump_off) | Peak2(pump_off) | Peak3(pump_off) | Peak4(pump_off) | Peak5(pump_off) |
| y0 | 0 ± 9.29981 | 0 ± 9.29981 | 0 ± 9.29981 | 0 ± 9.29981 | 0 ± 9.29981 |
| Xc | 0.00319 ± 0.00144 | 0.43159 ± 0.03698 | 0.73034 ± 0.00468 | 1.2444 ± 0.03129 | 1.61262 ± 0.10289 |
| W | 0.15582 ± 0.00449 | 0.15624 ± 0.13845 | 0.32711 ± 0.0222 | 0.44682 ± 0.14708 | 0.31234 ± 0.35447 |
| A | 610.70445 ± 13.66038 | 26.80123 ± 24.71401 | 779.89321 ± 57.77565 | 244.27916 ± 95.96047 | 37.35839 ± 58.91013 |
| Reduced Chi-Sqr | 3818.0421 | | | | |
| R-Square (COD) | 0.98709 | | | | |
| Adj. R-Square | 0.98539 | | | | |

Table S2. Peak fitting of pumped spectrum

| Model | Lorentz | | | | |
|---|---|---|---|---|---|
| Equation | y = y0 + (2*A/pi)*(w/(4*(x-xc)^2 + w^2)) | | | | |
| Plot | Peak1(pump_on) | Peak2(pump_on) | Peak3(pump_on) | Peak4(pump_on) | Peak5(pump_on) |
| y0 | 0 ± 5.31005 | 0 ± 5.31005 | 0 ± 5.31005 | 0 ± 5.31005 | 0 ± 5.31005 |
| Xc | 0.00319 ± 0 | 0.45257 ± 0.03811 | 0.7323 ± 0.00354 | 1.25464 ± 0.02159 | 1.57746 ± 0.06609 |
| W | 0.15497 ± 0.00276 | 0.17237 ± 0.07631 | 0.32881 ± 0.0071 | 0.48624 ± 0.04343 | 0.29889 ± 0.13246 |
| A | 610.70445 ± 0 | 26.80123 ± 0 | 779.89321 ± 0 | 244.27916 ± 0 | 37.35839 ± 0 |
| Reduced Chi-Sqr | 3623.08286 | | | | |
| R-Square (COD) | 0.9872 | | | | |
| Adj. R-Square | 0.98624 | | | | |

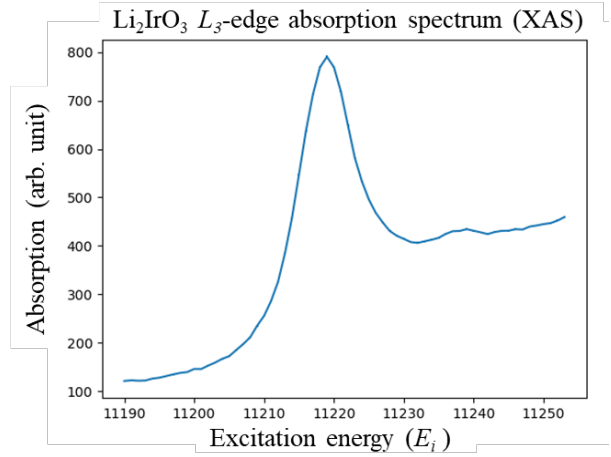

Figure. S1. Ir $L_3$-edge absorption spectrum of the studied α-Li$_2$IrO$_3$ single crystal sample.

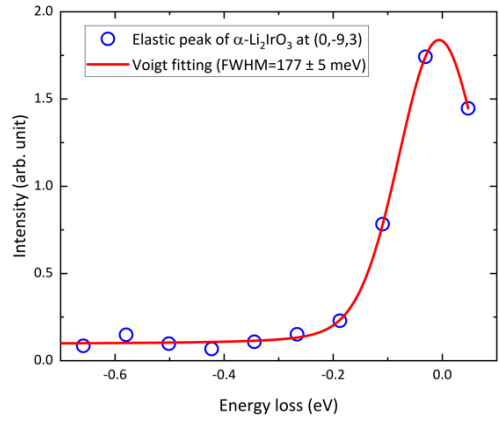

Figure. S2. Voigt fit of elastic peak.

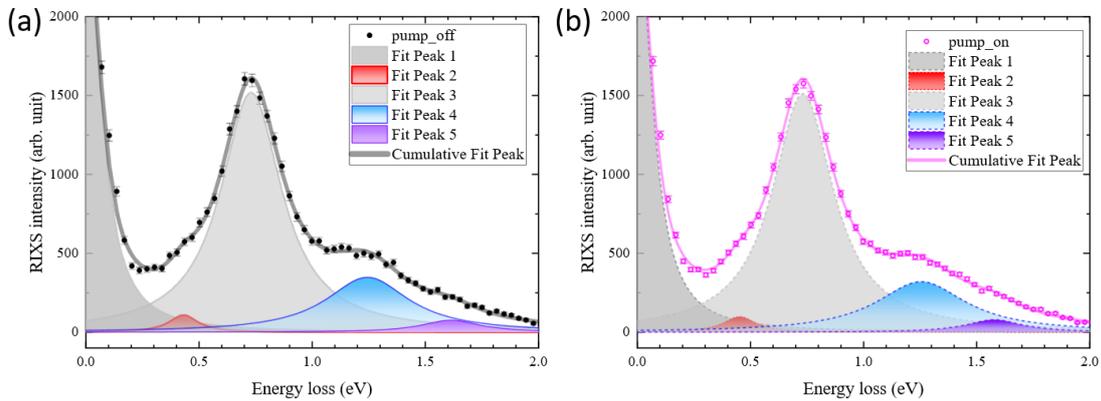

Figure S3: Fit of the pump-off (a) and pump-on (b) energy loss spectrum in the 0.2 eV region using five Lorentzian peaks including the elastic peak at 0 eV, the spin-orbit excitation at 0.7 eV, and itinerant excitations at 0.4 eV, 1.3 eV, 1.6 eV. Parameters are given in tables S1 and S2.

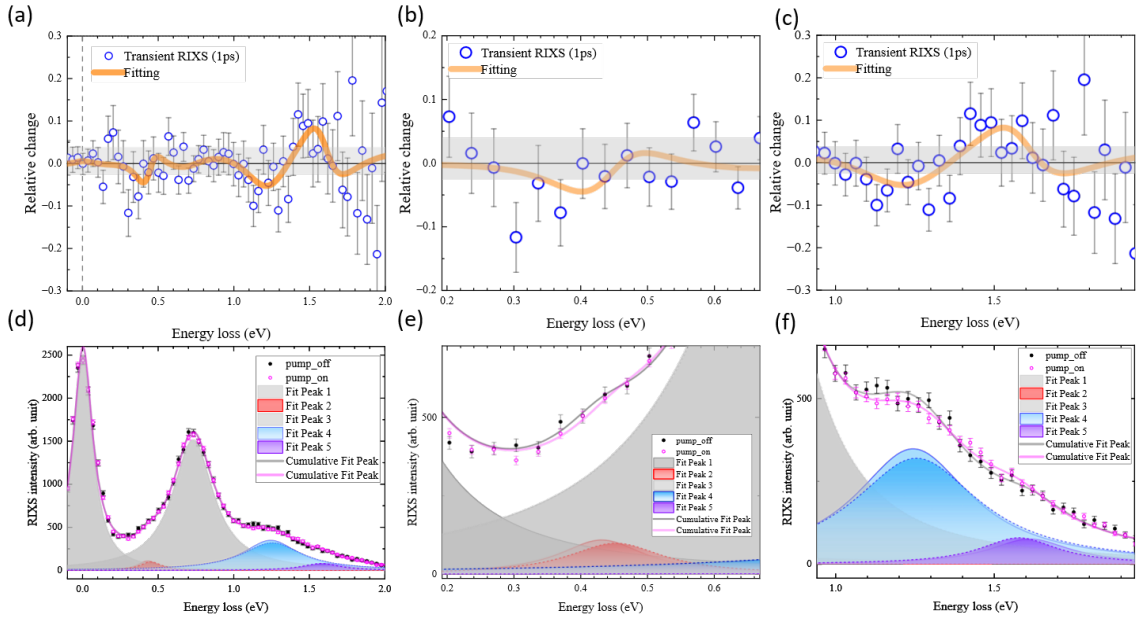

Figure S4. Transient RIXS spectra and its fitting with zooms-into two regions. (a) and (d) show all the fitted peaks, while (b), (e) are zooms-in of the 0.4 eV region, and (c), (f) are zooms-in of the 1.3~1.6 eV region.

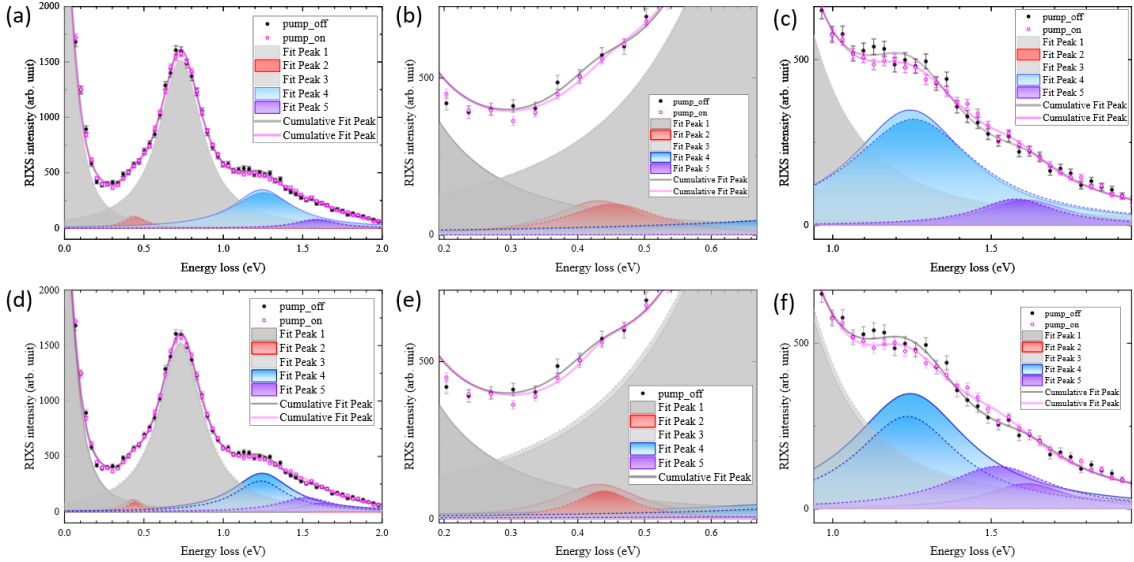

Figure S5: Comparison of two fittings for the pumped spectrum. (a) (b) (c) are fitted with area parameter fixed, where (b) and (c) are zoom-in for the energy loss region ~0.4 and 1.3eV respectively. (d)(e)(f) are fitted without constraints, where (e) and (f) are zoom-in for the energy loss region ~0.4 and 1.3eV respectively.

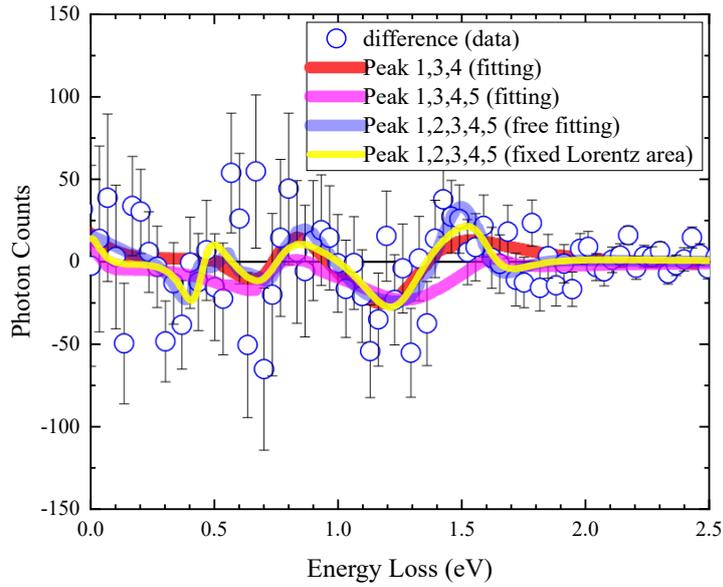

Figure S6: Comparison of fits of the RIXS transient (blue circles, pumped minus unpumped spectrum), considering different numbers of peaks. The fit traces using different combination of peaks in fitting are shown in different colors. The inclusion of peak 2 shows improved agreement with the transient spectrum, when comparing purple/yellow lines with the red/magenta lines.